\documentclass[twocolumn,showpacs,amsmath,amssymb,superscriptaddress,aps,prc,floatfix]{revtex4-1}
\usepackage[dvips]{graphicx}
\usepackage{epsfig}
\usepackage{amsmath}
\usepackage[dvips]{color}
\newcommand{\beq}{\begin{eqnarray}}
\newcommand{\eeq}{\end{eqnarray}}

\newcommand{\rmin}{\mbox{$r_{\rm min}$}}

\newcommand{\odp}{\mbox{$^{16}{\rm O}(d, p)^{17}$O}}

\newcommand{\clt}{\mbox{$^{12}{\rm C}(^7{\rm Li}, t)^{16}$O}}
\newcommand{\cago}{\mbox{$^{12}{\rm C}(\alpha,\gamma)^{16}$O}}
\newcommand{\can}{\mbox{$^{13}{\rm C}(\alpha,n)^{17}$O}}

\begin{document}
\title{Transfer reactions with the Lagrange-mesh method}
\author{Shubhchintak}
\email{khajuria1986@gmail.com}
\author{P. Descouvemont}
\email{pdesc@ulb.ac.be}
\affiliation{Physique Nucl\'eaire Th\'eorique et Physique Math\'ematique, C.P. 229,
Universit\'e Libre de Bruxelles (ULB), B 1050 Brussels, Belgium}
\date{\today}
\begin{abstract}
We apply the $R$-matrix method in Distorted Wave Born Approximation (DWBA) calculations. The internal wave
functions are expanded over a Lagrange mesh, which provides an efficient and fast technique to compute
matrix elements. We first present an outline of the theory, by emphasizing the $R$-matrix aspects. The model
is applied to the $\odp$ and $\clt$ reactions, typical of nucleon and of $\alpha$ transfer, respectively.
We illustrate the sensitivity of the cross sections with respect to the $R$-matrix parameters, and show
that an excellent convergence can be achieved with relatively small bases. We also discuss the effects of
the remnant term in DWBA calculations, and address the question of the peripherality in transfer reactions.
We suggest that uncertainties on spectroscopic factors could be underestimated in the literature.
\end{abstract}
\maketitle

\section{Introduction}
Transfer reactions represent an important tool to investigate the nuclear structure \cite{Sa83,Gl04,OII69,Ta74,Th88}. The
cross sections are known to be very sensitive to the structure of the projectile and of the residual
nucleus. For example, $(d,p)$ reactions have been widely used to probe the structure of many nuclei. Various 
models have been developed since more than 50 years (see Refs.\ \cite{GM14,Jo14} for recent reviews).
The angular distributions of a $A(d,p)B$ reaction permits identifying the angular momentum and the spectroscopic
factor of nucleus $B$. In particular, the development of radioactive beams led to many studies of exotic
nuclei (see, for example, Ref.\ \cite{CGM17}).

Transfer reactions are also commonly used in nuclear astrophysics as an indirect tool \cite{TBC14}. As radiative-capture 
cross sections are extremely small
at stellar energies, indirect measurements provide useful information on bound states and on low-energy resonances. Alpha-transfer
reactions, such as $(^6{\rm Li},d)$ or  $(^7{\rm Li},t)$, have been used to populate various states of $^{16}$O
\cite{BGK99} or of $^{17}$O \cite{PHR08}. These measurements are helpful to constrain the $\cago$ and $\can$ cross sections.
In parallel, nucleon-transfer reactions provide the
Asymptotic Normalization Constants (ANC) of several nuclei. For example, the authors of Ref.\ \cite{BBK00} 
perform a $^{13}{\rm C}(^{3}{\rm He},d)^{14}{\rm N}$ experiment to analyze $^{14}$N states. The deduced
ANCs are then used to determine the $^{13}{\rm C}(p,\gamma)^{14}{\rm N}$ cross section at low energies.

The theory of the Distorted Wave Born Approximation (DWBA) represents a standard framework for direct transfer
reactions \cite{ADH64}. In this approximation, the transition amplitude is determined as a first-order matrix element
of the transition potential between the initial and final scattering states. Various improvements, such as
the Coupled-Channel Born approximation (CCBA) \cite{PS64,CV76} have been proposed to include intermediate states in inelastic channels.

The calculation of transfer cross sections involves a three-body model for the entrance channel. In the DWBA,
the three-body wave function is factorized as a product of the target and projectile wave functions. Most transfer
calculations have been performed in this framework. The adiabatic method goes beyond this approximation by using
a genuine three-body wave function, but by assuming that the projectile is frozen during the collision \cite{JS70}.
More recently, the treatment of the three-body wave function has been improved by using the Faddeev method
\cite{DRN16} or the Coupled-Channel Discretized Continuum (CDCC) method \cite{MNJ09}.

Modern calculations, such as those involved in the CDCC method, are demanding in terms of computer capabilities. The
availability of efficient numerical techniques is therefore an important issue. Our goal in the present work is to apply 
the $R$-matrix method \cite{DB10,De16a} combined with the Lagrange-mesh theory \cite{Ba15} to transfer reactions.
In the $R$-matrix method, the configurations space is divided into two regions, separated by the channel radius $a$.
In the internal region, the wave functions are expanded over a basis. In the external region, the nuclear potential,
and the scattering wave functions have reached their asymptotic Coulomb behaviour, and the matching provides the scattering matrices
and the cross sections. Lagrange meshes correspond to specific bases, associated with the Gauss quadrature, and
have been applied in various fields of physics (see a review in Ref.\ \cite{Ba15}). When the matrix elements are computed at
the Gauss approximation, their calculation is greatly simplified, since numerical quadratures are not necessary. 

The paper is organized as follows. In Sec.\ \ref{sec2}, we briefly present the DWBA formalism, and emphasize on the 
$R$-matrix approach of transfer reactions. In Sec.\ \ref{sec3}, we apply the method to two examples.
The $\odp$ reaction is typical of nucleon transfer, whereas
the $\clt$ reaction is typical of $\alpha$ transfer. The conclusion and outlook
are presented in  Sec.\ \ref{sec4}

\section{Transfer reactions in the $R$-matrix formalism}
\label{sec2}
\subsection{Outline of the DWBA}
The DWBA theory has been presented in many reviews and books (see for example, Refs.\ \cite{Sa83,Gl04,OII69,Ta74,Th88}).  Here, we briefly introduce the method and define the notations.

Let us consider the rearrangement reaction
\beq
A(=a+v)+b \longrightarrow a+B(=b+v),
\label{eq1}
\eeq
where a valence cluster $v$ is transferred from the projectile $A$ to the target $b$.  A typical reaction is $(d,p)$ 
where a neutron $(v)$ is transferred, and a proton $(a)$ is emitted.  Pick-up processes (e.g. $(p,d)$ reactions) are described by a similar formalism.

\begin{figure}[htb]
	\begin{center}
		\epsfig{file=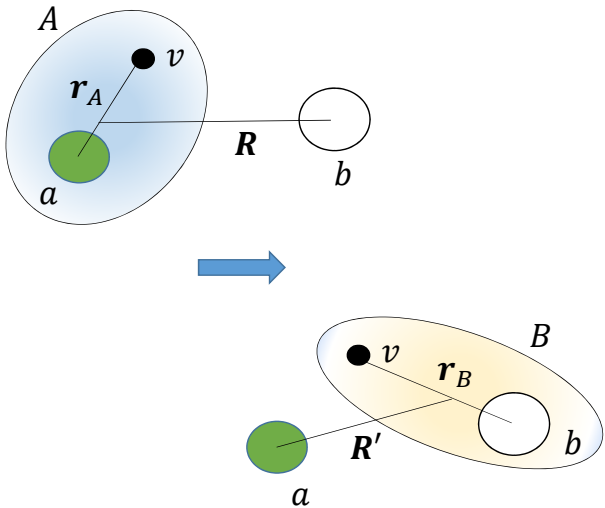,width=8cm}
		\caption{Schematic diagram of process (\ref{eq1}). The valence cluster $v$ is transferred from the
			projectile $A$ to the target $b$. The angles $\Omega$ and $\Omega'$ are associated with the coordinates
		$\pmb{R}$ and $\pmb{R}'$.}
		\label{fig1}
	\end{center}
\end{figure}

The various coordinates involved in the reaction (\ref{eq1}) are displayed in Fig.\ \ref{fig1}.  
Nuclei $A$ and $B$ are described in a two-body model, where the Hamiltonian is given by
\begin{align}
&H_A = T_A+V_{av}(\pmb{r}_A), \nonumber \\
&H_B=T_B+V_{bv}(\pmb{r}_B).
\label{eq2}
\end{align}
Potentials $V_{av}$ and $V_{bv}$ are real, and are fitted on spectroscopic properties of 
nuclei $A$ and $B$, such as binding energies, 
or rms radii.  In the final nucleus $B$, these potentials in general depend on the considered state.  
The two-body (bound-state) wave functions 
with spins $I_A$ and $I_B$ are written as
\begin{align}
\Phi^{I_AM_A}_{\ell_A}(\pmb{r}_A) =& \frac{1}{r_A} u^{I_A}_{\ell_A}(r_A)
\biggl[Y_{\ell_A}(\Omega_A)\otimes \bigl[\chi_a \otimes \chi_v\bigr]^{S_A} \biggr]^{I_AM_A} , \nonumber \\
\Phi^{I_BM_B}_{\ell_B}(\pmb{r}_B) =& \frac{1}{r_B} u^{I_B}_{\ell_B}(r_B)
\bigl[Y_{\ell_B}(\Omega_B)\otimes \chi_v\bigr]^{I_BM_B}  ,
\label{eq3}
\end{align}
where $\ell_A$ and $\ell_B$ are the orbital angular momenta (the parity is implied). 
In these definitions, $\chi_a$ and $\chi_v$ are spinors
associated with particles $a$ and $v$, respectively. We assume that the target nucleus $b$ has a spin zero, and cannot be excited.  A generalization can be found in Ref.\ \cite{GMG15}. The antisymmetrization between clusters $a$ and $v$ in the projectile, or
between $b$ and $v$ in the residual nucleus, is simulated by an appropriate choice of the potential, involving Pauli
forbidden states \cite{Fr81}. This technique uses deep potentials where the Pauli forbidden states are approximated by deeply bound states.

The natural sets of independent variables are $(\pmb{R},\pmb{r}_A)$ or $(\pmb{R}',\pmb{r}_B) $.  However, 
for symmetry reasons, the set $(\pmb{R},\pmb{R}') $ is usually adopted.  Then, the Jacobian ${\cal J}$ must be introduced in 
the matrix elements.  In the Appendix, we give more detail about the Jacobian and about the relationships 
between coordinates $(\pmb{r}_A,\pmb{r}_B,\pmb{r}_{ab}) $ and coordinates $(\pmb{R},\pmb{R}') $.

\subsection{Transfer scattering matrices}
The three-body Hamiltonian associated with reaction (\ref{eq1}) can be defined in the ``prior" representation \cite{Sa83,Gl04} as
\begin{align}
H_{\rm prior}=H_A(\pmb{r}_A)+T_{\pmb{R}}+V_{bv}(\pmb{r}_B)+V_{ab}(\pmb{r}_{ab}),
\label{eq4}
\end{align}
or, in the ``post" representation, as
\begin{align}
H_{\rm post}=H_B(\pmb{r}_B)+T_{\pmb{R}'}+V_{av}(\pmb{r}_A)+V_{ab}(\pmb{r}_{ab}),
\label{eq5}
\end{align}
where $\pmb{r}_{ab}$ is the distance between particles $a$ and $b$, and where $V_{ij}$ are optical potentials between 
the clusters.  In general, they are fitted on elastic-scattering 
data.  These two representations are strictly equivalent.  The merits of both choices are discussed, for example, 
in Refs.\ \cite{MNJ09,LM18}.  In the following, we use the post representation, but the developments are 
similar for the prior representation.  We assume here that all optical potentials are local. Extensions to
non-local potentials have been developed, for example, in Ref.\ \cite{WT16}.

Let us consider an auxiliary potential $V_{\beta}(\pmb{R}')$ between nuclei $a$ and $B$ in the exit channel.  
The corresponding three-body Hamiltonian is given by
\begin{align}
H_{\beta}=H_B(\pmb{r}_B)+T_{\pmb{R}'}+V_{\beta}(\pmb{R}'),
\label{eq6}
\end{align}
and the wave function with total angular momentum $J$ and parity $\pi$  can be factorized as
\begin{align}
\label{eq7}
&\Phi^{JM\pi(-)}_{\beta}(\pmb{r}_B,\pmb{R}')= \\
&\hspace{1cm} \frac{1}{R'} \chi^{J\pi}_{L_B}(R') \biggl[Y_{L_B}(\Omega')\otimes \bigl[\Phi^{I_B}_{\ell_B}(\pmb{r}_B)\otimes \chi_a\bigr]^{S_B}\biggr]^{JM}  ,\nonumber
\end{align}
where $L_B$ is the orbital momentum in the exit channel.

The scattering matrix between the initial and final states can be written, for any choice of the auxiliary potential, as
\begin{align}
U^{J\pi}_{\alpha \beta}= -\frac{i}{\hbar}\langle \Phi^{JM\pi(-)}_{\beta} \vert
V_{av}+V_{ab}-V_{\beta}\vert  \Psi^{JM\pi(+)}_{\alpha}\rangle,
\label{eq8}                                 
\end{align}                                 
where $\Psi^{JM\pi(+)}_{\alpha}$ is the exact solution of the three-body equation (see Eq.\ (8.52) of Ref.\ 
\cite{Gl04}).  In 
definition (\ref{eq8}), we use the labels $\alpha=(L_A,\ell_A,I_A)$ and $\beta=(L_B,\ell_B,I_B)$.  At the DWBA, the exact three-body wave function $\Psi^{JM\pi(+)}_{\alpha}$ is replaced by
\begin{align}
\Psi^{JM\pi(+)}_{\alpha}(\pmb{r}_A,\pmb{R})\approx\frac{1}{R} \chi^{J\pi}_{L_A}(R)
\bigl[Y_{L_A}(\Omega)\otimes \Phi^{I_A}_{\ell_A}(\pmb{r}_A)\bigr]^{JM},
\label{eq9}
\end{align}
where $\chi^{J\pi}_{L_A}(R)$ is generated by an $A+b$ optical potential.  The two wave functions 
(\ref{eq7}) and (\ref{eq9}) 
are therefore treated on an equal footing.  

Notice that, owing to the use of the DWBA, the choice of the 
auxiliary potential $V_{\beta}$ is not arbitrary (see the discussion in Ref.\ \cite{MNJ09}).  A common choice is 
to adjust this potential on elastic-scattering data, which ensures the correct asymptotic behaviour of the
scattering wave function.  An alternative consists in using a folding potential from $V_{av}+V_{ab}$.  This option 
is more consistent in the sense that it is based on the three-body Hamiltonian (\ref{eq5}), without any 
additional input. However, this folding potential may not be optimal for elastic scattering.

There are various approaches that go beyond the DWBA.  In the adiabatic approximation \cite{JS70,Jo14}, 
the three-body wave function is approximated as
\begin{align}
\Psi^{JM\pi(+)}_{\alpha}(\pmb{r}_A,\pmb{R})\approx\frac{1}{R} \chi^{J\pi}_{L_A}(R,r_A)
\bigl[Y_{L_A}(\Omega)\otimes \Phi^{I_A}_{\ell_A}(\pmb{r}_A)\bigr]^{JM} ,
\label{eq10}
\end{align}
and assumes that the projectile $A$ is ``frozen" during the collision.  This approximation is valid when 
the scattering energy is much higher than the binding energy of the projectile.  It has been extensively 
used for $(d,p)$ reactions \cite{Jo14}.

The Continuum Discretized Coupled Channel (CDCC) model \cite{YIK86,AIK87,YOM12} aims at improving the three-body 
wave function (\ref{eq9}) by discretizing the continuum of nucleus $A$ over a set of square-integrable two-body wave 
functions $\Phi^{I_A}_{\ell_A,n}(\pmb{r}_A)$ as
\begin{align}
\label{eq11}
\Psi^{JM\pi(+)}_{\alpha}(\pmb{r}_A,\pmb{R})\approx\frac{1}{R}&\sum_n  \chi^{J\pi}_{L_A,n}(R)  \\
&\times \bigl[Y_{L_A}(\Omega)\otimes \Phi^{I_A}_{\ell_A,n}(\pmb{r}_A)\bigr]^{JM},\nonumber
\end{align}
where index $n$ denotes, either bound states, or approximate continuum states.
Here the radial functions $\chi^{J}_{L_A,n}(R)$ are obtained from a coupled-channel system. The CDCC 
method has been used for many reactions involving weakly bound nuclei, where breakup effects are expected 
to be important.  The use of CDCC wave functions in transfer reactions is more recent \cite{MNJ09,GM17}.

In Eq.\ (\ref{eq8}), the core-core potential and the auxiliary potential $V_{\beta}$ are similar.  Then, 
the remnant potential 
\begin{align}
V_{\rm rem}=V_{ab}-V_{\beta}
\label{eq12}
\end{align}
is often neglected.  This approximation is expected to be valid for a nucleon transfer on a heavy target.  It is 
reasonable to assume that the $a+b$ potential is close to the $a+(b+1)$ potential if the target $b$ is heavy.  
For light targets, however, and for $\alpha$ transfer, remnant effects may be not negligible.

In the DWBA, the scattering-matrix element (\ref{eq8}) is given by
\begin{align}
U^{J\pi}_{\alpha \beta}= -\frac{i}{\hbar}\int 
\chi^{J\pi}_{L_A}(R) K^{J\pi}_{\alpha \beta}(R,R') \chi^{J\pi}_{L_B}(R')RR'dRdR',
\label{eq13}
\end{align}
where the $ K^{J\pi}_{\alpha \beta}(R,R')$ transfer kernel is defined as
\begin{align}
&K^{J\pi}_{\alpha \beta}(R,R') ={\cal J}\langle 
\bigl[Y_{L_A}(\Omega)\otimes \Phi^{I_A}_{\ell_A}(\pmb{r}_A)\bigr]^{J} 
\vert V_{av}+V_{\rm rem}\vert \nonumber \\
&\hspace{1cm} \vert \bigl[Y_{L_B}(\Omega')\otimes \Phi^{I_B}_{\ell_B}(\pmb{r}_B)\bigr]^{J} 
\rangle .
\label{eq14}
\end{align}
The calculation is developed in the Appendix.  It involves integrals over the angles $\Omega$ and 
$\Omega' $.  Notice that the integral definitions (\ref{eq8},\ref{eq13}) of the scattering matrix assume 
that the scattering wave functions tend to
\begin{align}
\chi_L(r)\rightarrow\frac{1}{\sqrt{v}}(I_L(kr)-U_L O_L(kr)),
\label{eq15}
\end{align}
where $k$ and $v$ are the wave number and velocity, respectively.  This definition also involves the incoming 
and outgoing Coulomb functions $I_L(x)$ and $O_L(x)$, as well as the elastic scattering matrix $U_L$.  
The calculation of $U_L$ and of $\chi_L(r)$ is further discussed in the next subsection.
Definition (\ref{eq13}) can be easily extended to CDCC
wave functions (\ref{eq11}) by including additional summations over the different channels.

When the scattering matrices (\ref{eq13}) are known, the transfer cross sections can be computed 
(see, for example, Ref.\ \cite{DB10}). The integrated transfer cross section is given by 
\begin{align}
	\sigma_t=\frac{\pi}{k^2 (2I_A+1)}\sum_J (2J+1) T_J,
\label{eq15b}
\end{align}
with
\begin{align}
 T_J=\sum_{\pi}\sum_{L_A,I_B,L_B}\vert U^{J\pi}_{I_AL_A,I_BL_B}\vert ^2.
\label{eq15c}
\end{align}

\subsection{The $R$-matrix method}
	
The definition of the transfer scattering matrix (\ref{eq13}) is general.  Besides the $K^{J\pi}_{\alpha \beta}$ transfer kernel, 
these integrals involve scattering wave functions $\chi^{J\pi}_{L_A}(R)$ and $\chi^{J\pi}_{L_B}(R')$.  In the 
FRESCO code \cite{Th88} they are obtained with a finite-difference method.  This discretization method, however,
usually requires many points to get a good  accuracy.  In simple calculations, the computer time is always short, 
and does not represent an important issue.  In more complex calculations, such as those using the CDCC method 
(see, for example, Ref.\ \cite{De17} for a recent application), a special attention must be paid to the numerical procedure.
	
In the present work, we use the $R$-matrix method \cite{LT58,DB10,De16a} to determine the scattering wave 
functions $\chi^{J\pi}_{L_A}(R)$ and $\chi^{J\pi}_{L_B}(R')$.  Although we limit this short presentation to
single-channel problems, the formalism can be easily extended to multichannel problems \cite{DB10}, such as 
those encountered in CDCC calculations.  

The basic idea of the $R$-matrix theory is to divide the space in 
an internal region (with radius $a$) and in an external region.  The channel radius $a$ should be large 
enough so that the nuclear potential is negligible.  In the internal region $(R\leq a)$, the wave function is 
expanded over a set of $N$ basis functions $\varphi_i(R)$ as
\begin{align}
\chi^L_{\rm int}(R)=\sum_{i=1}^N c^L_i \varphi_i(R),
\label{eq16}
\end{align}
where the choice of functions $\varphi_i(R)$ will be discussed later (in this subsection, we only write the
relative orbital momentum $L$ for the sake of clarity).  In the external region $(R > a)$, by definition 
of the channel radius, the wave function takes the asymptotic form (\ref{eq15}) as
\begin{align}
\chi^L_{\rm ext}(R)=\frac{1}{\sqrt{v}}(I_L(kR)-U_L O_L(kR)),
\label{eq17}
\end{align}
where $U_L$ is the scattering matrix for elastic scattering and is a number for single-channel 
problems. Since the basis functions $\varphi_i(R)$ are valid for $R\leq a$ only, matrix elements of the 
kinetic energy are not Hermitian.  This is addressed by introducing the Bloch operator
\begin{align}
{\cal L}=\frac{\hbar^2}{2\mu}\delta(R-a)\biggl( \frac{d}{dR}-\frac{B}{R}\biggr),
\label{eq18}
\end{align}
where $\mu$ is the reduced mass, and $B$ is a boundary parameter, taken here as $B=0$.  
The role of the Bloch operator is twofold: it 
ensures the hermiticity of the Hamiltonian over the internal region, and the continuity of the derivative 
at the surface.  Then the Bloch-Schr\"{o}dinger equation reads
\begin{align}
\bigl(H+{\cal L}-E \bigr)\chi^L_{\rm int}={\cal L} \chi^L_{\rm int}={\cal L} \chi^L_{\rm ext},
\label{eq19}
\end{align}
where the second equality holds from the surface character of ${\cal L}$.  

Inserting the expansion (\ref{eq16}) in (\ref{eq19}) provides coefficients $c^L_i$ as
\begin{align}
c^L_i=\sum_j (C_L^{-1})_{ij}\langle \varphi_j \vert {\cal L}\vert \chi^L_{\rm ext}\rangle ,
\label{eq20}
\end{align}
where matrix $\pmb{C}_L$ is defined by
\begin{align}
\bigl(\pmb{C}_L\bigr)_{ij}=\langle \varphi_i \vert H+{\cal L}-E \vert \varphi_j \rangle.
\end{align}
The continuity condition
\begin{align}
\chi^L_{\rm int}(a)=\chi^L_{\rm ext}(a)
\label{eq21}
\end{align}
gives the scattering matrix
\begin{align}
U_L=\frac{I_L(ka)}{O_L(ka)} \frac{1-L^{\ast}R_L}{1-LR_L},
\label{eq22}
\end{align}
where constant $L$ is defined as
\begin{align}
L=S_L+iP_L=ka \frac{O'_L(ka)}{O_L(ka)}.
\label{eq23}
\end{align}
The real part $S_L(E)$ and the imaginary part $P_L(E)$ are known as the shift and penetration functions, respectively.
In Eq.\ (\ref{eq22}), the $R$-matrix is obtained from
\begin{align}
R_L=\frac{\hbar^2}{2\mu a}\sum_{i j}\varphi_i(a)
\bigl( \pmb{C}_L^{-1}\bigr)_{ij}\varphi_{j}(a).
\label{eq24}
\end{align}
	
From the $R$-matrix, the elastic scattering matrix $U_L$ as well as coefficients $c^L_i$ are easily determined.  Let us 
point out that the channel radius is not a parameter.  Although the $R$ matrix and the Coulomb functions 
in (\ref{eq22}) do depend on $a$, the scattering matrix should not depend on it.  The choice of the channel 
radius results from a compromise: on the one hand, it should be large enough to make sure that the nuclear 
interaction is negligible.  On the other hand, large values require a large number of basis functions 
$\varphi_i(R)$ which increases the computer times.  In this respect a channel radius as small as possible
is recommended. In the next subsection we discuss Lagrange functions, 
which represent an efficient choice in $R$-matrix calculations.
	
In practice, the main part of the computation time comes from the inversion of the complex matrix $\pmb{C}_L$ [Eq.\
(\ref{eq21})]. When the channel radius $a$ is large or, in other words, when the nuclear potential extends to large
distances, the corresponding number of points must be increased.
This issue can be addressed by using propagation techniques \cite{BBM82,DB10}, where the interval $[0,a]$
is split in subintervals. These techniques allow to deal with large numbers of coupled equations, since the number
of Lagrange functions can be reduced in each subinterval. Consequently the sizes of the matrices to be inverted are smaller.

\subsection{The DWBA method with Lagrange meshes}
	
The Lagrange functions have been used in different fields of physics \cite{Ba15}.  The main idea is to define basis functions $ $ associated with a Gauss quadrature.  The functions depend on the interval considered.  For a finite 
interval $[0,a]$, such as those encountered in the $R$-matrix theory, the $N$ Lagrange functions are chosen as
\begin{align}
\varphi_i (R)=(-1)^{N+i} \sqrt{\frac{x_i(1-x_i)}{a x_i}}  \,
\frac{R\,P_N(R/a-1) }{R-a x_i},
\label{eq25}
\end{align}
where $P_N(x)$ is the Legendre polynomial of degree $N$, and $x_i$ are the zeros of
\beq
P_N(2x_i-1)=0.
\label{eq26}
\eeq
These functions satisfy the Lagrange conditions
\begin{align}
\varphi_i(ax_j)=\frac{1}{\sqrt{a\lambda_i}}\delta_{ij},
\label{eq27}
\end{align}
where $\lambda_i$ are the weights of the Gauss-Legendre quadrature in the $[0,1]$ interval.  This mesh is used
for the scattering wave functions $\chi^{J\pi}_{L_A}(R)$ and $\chi^{J\pi}_{L_B}(R')$.

For bound states, the interval ranges from $0$ to infinity, and the Lagrange functions are associated with 
Laguerre polynomials $L_N(x)$ as
\begin{align}
\varphi_i(R)=(-1)^i \frac{R}{R-x_ih}\frac{1}{\sqrt{x_i}}L_N(R/h)\exp(-R/2h),
\label{eq28}
\end{align}
where $h$ is a scale parameter, adapted to the typical dimensions of the system.  These basis functions are used
to describe the bound states of nuclei $A$ and $B$.
In Eq.\ (\ref{eq28}), the $x_i$ are 
the roots of the Laguerre polynomials of order $N$.  The Lagrange condition reads
\begin{align}
\varphi_i(h\,x_j)=\frac{1}{\sqrt{h\lambda_i}}\delta_{ij},
\label{eq29}
\end{align}	
where $\lambda_i$ are now the weights associated with the Gauss-Laguerre quadrature.
	
Lagrange functions are very efficient when the matrix elements are computed consistently at the Gauss quadrature 
of order $N$.  Matrix elements of the overlap and of a local potential $V(r)$ are given by
\begin{eqnarray}
\langle\varphi_i \vert \varphi_j \rangle&\approx&\delta_{ij} \nonumber \\
\langle\varphi_i \vert V \vert \varphi_j \rangle&\approx&V(ax_i)\delta_{ij}{\rm \ for\ Legendre\ functions}\nonumber \\
&\approx&V(hx_i)\delta_{ij} {\rm \ for\ Laguerre\ functions}.
\label{eq30}
\end{eqnarray}
According to Eq.\ (\ref{eq30}), all matrix elements only require the values of the potential at the mesh points.    This property can be even extended to non-local potentials \cite{HRB02}. The accuracy of the Gauss approximation in the Lagrange-mesh method
has been discussed in the literature (see, for example, Refs.\ \cite{BHV02,Ba15}). The 
matrix elements of the kinetic energy can be found in Ref.\ \cite{Ba15}.
	
If the channel radius is chosen large enough so that the external contribution in Eq.\ (\ref{eq13}) is negligible,
the calculation of the transfer scattering matrix (\ref{eq13}) takes the simple form
\begin{align}
U^{J\pi}_{\alpha \beta}=- \frac{i}{\hbar}a^3 \sum_{i,j=1}^Nc^{\alpha}_i c^{\beta}_j 
\sqrt{\lambda_i \lambda_j} x_i x_j K^{J\pi}_{\alpha \beta}(ax_i,ax_j),
\label{eq31}
\end{align}
where coefficients $c^{\alpha}_i$ and $c^{\beta}_j$ are associated with the scattering wave 
functions $\chi^{J\pi}_{L_A}(R)$ and $\chi^{J\pi}_{L_B}(R')$, respectively.  Typically, $N\sim 30-40$ points are sufficient 
to achieve convergence. This is significantly less than in finite-difference methods, where typical
number of points is typically of the order of 500. Of course, the $R$-matrix method involves additional steps, such as the
inversion of a complex matrix (see the discussion in Sec.\ II.C), but this can be efficiently optimized.
	
Gauss-Laguerre functions are used to expand the bound-state radial functions $u_{\alpha}(r_A)$ and $u_{\beta}(r_B)$.  
This procedure involves fast calculations, and is efficient for large-scale calculations.  Of course the 
transfer cross sections should not depend on the numerical approach. As for elastic scattering, the transfer 
cross sections must be stable against variations of the channel radius and of the number of basis functions.
Several tests have been performed with the code FRESCO \cite{Th88}, and will be discussed in the next Section.
In optimal conditions (i.e. with numbers of points as small as possible), the Lagrange-mesh method is about two 
times faster than FRESCO.

\section{Results and discussions}
\label{sec3}
\subsection{Conditions of the calculations}
We consider two reactions: $\odp$ and $\clt$, involving the transfer of a neutron,  
and of an alpha particle, respectively. Although we compare our calculations with the available experimental 
data, we want to stress that our goal is not to fit the data or to extract spectroscopic factors (SFs). 
Our aim is to assess the use of the $R$-matrix method in transfer calculations.
In particular, the important parameters in the current approach are the number of basis functions $N$ 
and the channel radius $a$, which are chosen large enough to ensure converged results. Throughout the paper,
we use integer masses, and the constant $\hbar^2/2m_N=20.9$ MeV.fm$^2$ ($m_N$ is the nucleon mass). Unless specified otherwise, 
we assume that spectroscopic factors are unity.

Other important inputs are the bound and scattering state potentials required to generate the corresponding wave functions. For bound-state calculations, we use  potentials given in the literature,  which reproduce the binding energy of the concerned state. 
The deuteron ground-state wave function ($s$ state) is calculated with the standard Gaussian potential
\begin{align}
V_{np}(r) = -72.66\, \exp[-(r/1.484)^2].
\end{align}
The bound state potentials of the other systems are taken of the Woods-Saxon type, as
\begin{align}
\label{vb}
V(r) = &-V_r \,f(r, R_r, a_r) + V_c(r)\nonumber \\
& -  V_{so} \Big(\frac{\hbar}{m_\pi\,c}\Big)^2\frac{1}{r}\frac{d}{dr}f(r,R_{so},a_{so}) \pmb {\ell \cdot s}.
\end{align}
with
\begin{align}
f(r,R,a) = 1/[1+\exp(\frac{r-R}{a})].
\end{align}
Here, $V_c$ is the Coulomb potential of an uniformly charged sphere with radius $R_c$ and $m_{\pi}$ the pion mass. 
In comparison with the original references,
the amplitudes are slightly adjusted to reproduce the experimental binding energies with the adopted physical constants. The various parameters are given in Table \ref{tab:pot_b}. 

\begin{table}[ht]
	\caption{Woods-Saxon potential parameters for bound states.  }
	\label{tab:pot_b}
	\begin{ruledtabular}
		\begin{tabular}{ccccccccc}
			System &state    & $V_r$ & $R_r$  & $a_r$ & $V_{so}$ & $R_{so}$ & $a_{so}$ & $R_c$\\
			&        & (MeV) & (fm)   & (fm)  & (MeV) & (fm)  & (fm)  & (fm)   \\   
			\hline       
			$n+^{16}$O\tablenotemark[1] &$5/2^+$    &52.96 & 3.15 &  0.523  & 5.332 &  3.15 & 0.523 & \\
			&$1/2^+$    &54.97 & 3.15 &  0.523  & 5.33 &  3.34 & 0.523 &   \\
			$\alpha+t$\tablenotemark[2] & $3/2^-$     &94.0   &2.05 &  0.70    &       &        &       & 2.05\\
			$\alpha+^{12}$C\tablenotemark[2]& $0^+_2$&71.1  & 4.50 &  0.53 &       &       &       & 5.0\\
			&$2^+_1$ &69.15  & 4.50 &  0.53 &      &        &       & 5.0\\
		\end{tabular}
	\end{ruledtabular}
	\tablenotetext[1] {Ref.\ \cite{CHR74}.}
	\tablenotetext[2]{Ref.\ \cite{OHR12}.}
\end{table}

To calculate the scattering wave functions, we also make use of phenomenological optical potentials available in
the literature.  These potentials are obtained by fitting elastic-scattering data, and are of the form
\begin{align}
\label{ws1}
U(r) = &-V_r \,f(r, R_r, a_r)+V_c(r) \nonumber \\
&- i\,W_v\,f(r,R_v,a_v)-i\,W_s\,g(r,R_s,a_s). 
\end{align}
The imaginary potential contains a volume term  and a surface term defined by
\begin{eqnarray}
g(r,R_s,a_s) = -4\,a_s\,\frac{d}{dr}f(r,R_s,a_s). \label{ws2}
\end{eqnarray}
For $^{17}$O+p, a Gaussian surface imaginary \cite{CHR74} is used as
\begin{eqnarray}
g(r,R_s,a_s) = \exp \bigl( -\Big[\frac{0.69(r-R_s)}{a_s}\Big]^2\bigr). 
\label{ws3}
\end{eqnarray}
Table \ref{tab:pot1} contains the parameters of the various optical potentials. 
For the sake of simplicity, and since our goal is not to obtain best fits of the data, we neglect spin-orbit
effects.

\begin{table*}[t]
	\caption{Optical potential parameters defined by Eqs. (\ref{ws1}-\ref{ws3}), for the various channels involved in the reactions considered in this paper.  }
	\label{tab:pot1}
	\begin{ruledtabular}
		\begin{tabular}{lcccccccccccc}
			channel&  $E_{\rm lab}$  & $V_r$ & $R_r$  & $a_r$ & $W_v$  & $R_v$ & $a_v$ & $W_s$ & $R_s$ & $a_s$ &$R_c$ &Ref. \\
			&  (MeV)   & (MeV) & (fm)   & (fm)  & (MeV)  & (fm)  & (fm)  & (MeV) & (fm)  & (fm)  & (fm)  \\          
			\hline
			$d+^{16}$O      & 25.4     & 94.79 & 2.65   & 0.84 &        &       &       & 8.58  & 3.96 & 0.57 & 3.28 &  \cite{CHR74}\\
			& 36       & 92.84 & 2.59  & 0.80 &        &       &       & 8.84  & 3.55 & 0.697 & 3.28 & \cite{CHR74}\\ 
			\hline
			$p+^{17}$O$\tablenotemark[1]$    & 25.4     & 50.30 & 2.94 & 0.73 &        &       &       & 9.60  & 1.27 & 0.68 & 3.34 & \cite{CHR74}\\
			& 36       & 46.70 & 2.94  & 0.73 & 1.40   & 3.26 & 0.68 & 7.50  & 3.26 & 0.68 & 3.34 &\cite{CHR74}\\ 
			$p+^{17}$O(0.87)$\tablenotemark[1]$& 25.4& 50.70& 2.94  & 0.73 &        &       &       & 9.80  & 3.26 & 0.68 & 3.34 & \cite{CHR74}\\
			& 36       & 47    & 2.94  & 0.73 & 1.15   & 3.26 & 0.68 & 7.70  & 3.26 & 0.68 & 3.34& \cite{CHR74} \\        
			\hline
			$^7$Li+$^{12}$C& 28, 34&139.1& 3.71& 0.58 & 18.8   & 4.56  & 0.93  & 0     &    0  &    0  & 2.91& \cite{OHR12}\\
			\hline
			$t+^{16}$O& 28, 34&170  & 2.87   & 0.723 & 20     &  4.03  & 0.8   & 0     &    0  &    0  & 3.12 &\cite{OHR12}\\
			\hline
			$t+^{12}$C    & 28       &170.451& 2.41  & 0.73 & 13.85 &  2.85& 1.16 &19.00 & 2.40 & 0.84 & 3.26 &\cite{LLC07}\\
			& 34       &185.796& 2.41  & 0.73 & 13.255 &  2.85& 1.16 &15.13 & 2.40 & 0.84 & 3.26 &\cite{LLC07}\\
			\hline
			$t+^{12}$C    & 28       &138.48 & 2.29   & 0.72 & 2.49  &  3.11 & 0.80 &11.22 & 1.36  & 0.80 & 2.98 &\cite{PDM15}\\
			& 34       &134.41 & 2.31  & 0.792 & 2.76  &  3.11 & 0.80 & 10.92& 1.36  & 0.80 & 2.98 &\cite{PDM15}
		\end{tabular}
	\end{ruledtabular}
	\tablenotetext[1] {Gaussian surface imaginary potential.}
\end{table*}

\subsection{The $\odp$ reaction}
As a first example, we consider the stripping of a deuteron on $^{16}$O, leading to the ground ($5/2^+$) and to the first 
excited state ($1/2^+, E_x = 0.87$ MeV) of $^{17}$O at two different beam energies $E_d=25.4$ MeV and 36 MeV. 
The $^{17}$O states are constructed by coupling the $0^+$ ground state of $^{16}$O with a neutron in $1d_{5/2}$ and 
$2s_{1/2}$ orbitals, respectively. This reaction represents an excellent test case which has been abundantly covered in 
the literature (see, for example,
Ref.\ \cite{CHR74}). It has been reconsidered recently \cite{MGL19}.

\begin{figure}[ht]
	\centering
	\includegraphics[width=8.5cm,clip]{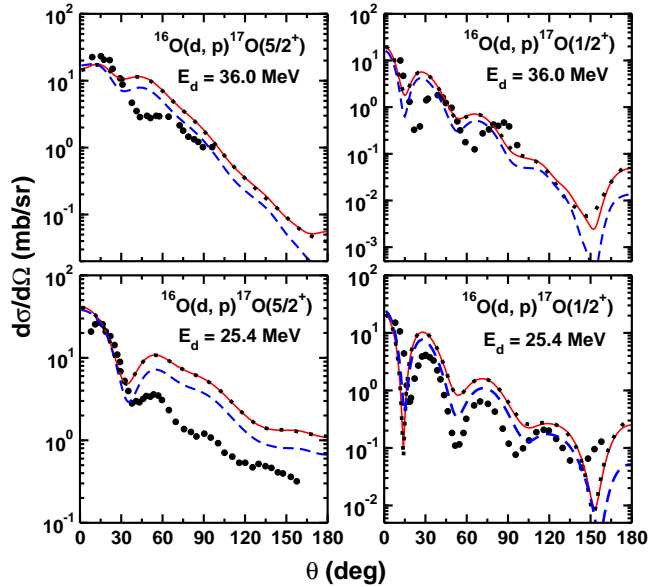}
	\caption{$\odp$ angular distributions for the ground ($5/2^+$) and first excited ($1/2^+$) states of $^{17}$O at two deuteron energies. Dashed and solid lines correspond to the calculations with and without the inclusion of the remnant term in the potential. The dotted lines correspond to the FRESCO calculations
		\cite{Th88}. Experimental data are taken from Ref.\ \cite{CHR74}. }
	\label{dp_css}
\end{figure}

In Fig.\ \ref{dp_css}, we plot the $\odp$ angular distributions and compare them with the experimental data 
of Ref.\ \cite{CHR74}.
From the figure one can see that at forward angles ($\lesssim 30^{\circ}$), the DWBA calculations are close to the data. Our results
are consistent with those of Ref.\ \cite{CHR74}. In that reference, the importance of breakup channels was discussed, 
which resulted in the improvement of the calculated cross sections at backward angles. Tests with the FRESCO code
(dotted lines) show an
excellent agreement with the present calculation.

We further investigate the importance of the remnant term in the $\odp$ reaction. In general, this
approximation greatly simplifies the calculations, and is often used in the literature. Going beyond this approximation, however, raises the question of a core-core potential. In the present work, we take the
$^{16}{\rm O}+p$ optical potential from the global parametrization of Ref.\ \cite{KD03}. Figure \ref{dp_css} 
shows that calculations performed without (solid lines) and with (dashed lines)
the remnant term are similar, especially at forward angles. At large angles, the difference may reach up to 30\%.

\begin{figure}[ht]
	\centering
	\includegraphics[width=7.5cm,clip]{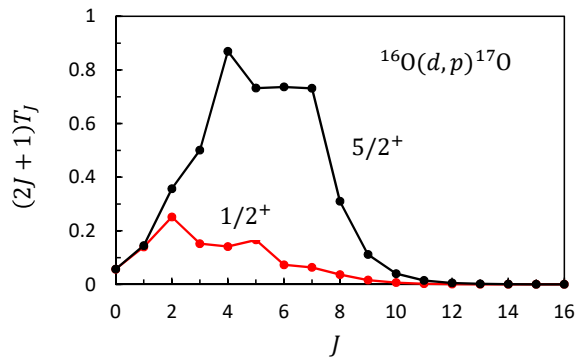}
	\caption{Coefficients $T_J$ (\ref{eq15c}) for the $\odp$ 
		reaction at $E_d =  25.4$ MeV as a function of $J$. The lines are guide to the eye. }
	\label{dp_J_odp}
\end{figure}

In Fig.\ \ref{dp_J_odp}, we plot the coefficients $(2J+1)T_J$ as a function of $J$  at $E_d=25.4$ MeV for 
the ground state as well as for the first excited state  of $^{17}$O. This quantity 
is relevant for the calculation of the integrated cross sections. In agreement with the cross sections of Fig.\ \ref{dp_css},
the $5/2^+$ contribution is larger.
Figure \ref{dp_J_odp} shows that partial waves $J \geq 10 $ have a small contribution. The maxima are located at low $J$ values ($J\approx 4-8$ for the ground state and $J\approx 2-5$ for the excited state). 

\begin{figure}[h]
	\centering
	\includegraphics[width=7.5cm,clip]{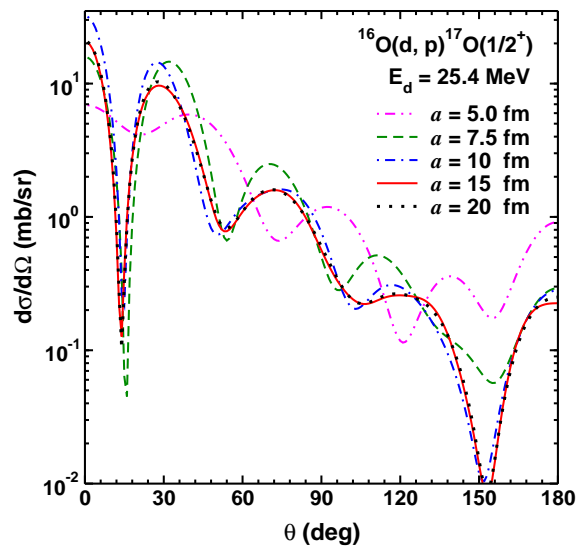}
	\caption{$\odp$ transfer cross section to the $1/2^+$ state at $E_d = 25.4$ MeV for various channel radii.  }
	\label{fig_conv1}
\end{figure}

\begin{figure}[h]
	\centering
	\includegraphics[width=7.5cm,clip]{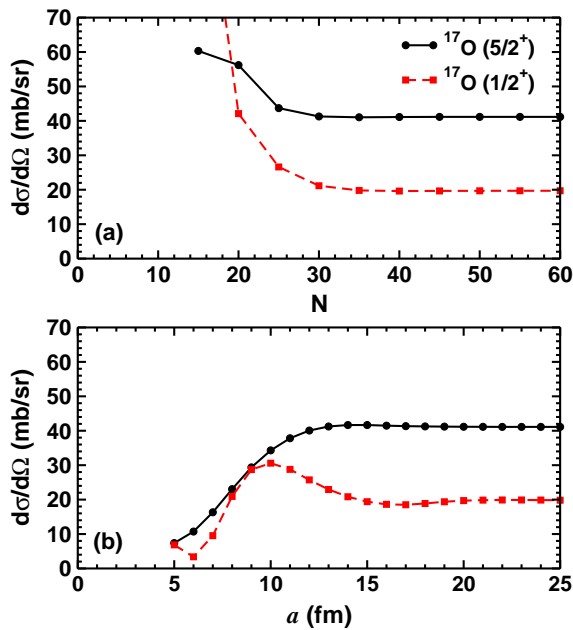}
	\caption{$\odp$ transfer cross section at $E_d = 25.4$ MeV and $\theta=2^{\circ}$ as
		a function of the number of basis functions $N$ (a) and of the channel radius (b). }
	\label{fig_conv2}
\end{figure}

In Figs.\ \ref{fig_conv1} and \ref{fig_conv2}, we analyze the sensitivity of the transfer cross sections against
variations of the $R$-matrix parameters: the channel radius $a$ and the number of basis functions $N$.
The channel radius must be large enough to guarantee that nuclear effects are negligible. However, large values
require large bases, and therefore increase the computer times. As usual in $R$-matrix calculations, a compromise must be adopted.

Figure \ref{fig_conv1} presents the $\odp$ cross section at $E_d=25.4$ MeV, and for various channel radii. The number of basis functions is fixed at a conservative value $N=80$. From the figure, we conclude that, as soon as the channel radius
is $a \gtrsim 15$ fm, the convergence is achieved. Similar conclusions are drawn at other energies.

In Fig.\ \ref{fig_conv2}, we select the scattering angle $\theta=2^{\circ}$, and plot the cross section for various $a$ and $N$. 
In Fig.\ \ref{fig_conv2}(a), we consider the variation of the cross section with the number of basis functions $N$.
Values around $N\approx 40$ are sufficient to achieve an excellent convergence. These numbers are much smaller than those use in finite-difference methods, such as the Numerov algorithm (several hundreds with a typical mesh size of
0.02 fm). As mentioned earlier, the $R$-matrix calculations can be still speed up by using a propagation 
method \cite{BBM82}.  This tool is particularly efficient in coupled-channel calculations involving many channels.
Figure \ref{fig_conv2}(b) confirms the previous analysis: a channel radius larger than $\sim 15$ fm is
necessary to ensure the convergence. Notice that the $^{17}{\rm O}(5/2^+)$ transfer cross section converges faster
than the $^{17}{\rm O}(1/2^+)$  cross section, due to the larger binding energy of the $5/2^+$ state. 

\subsection{The $\clt$ reaction}
In this subsection, we apply the formalism to the $\clt$ reaction, which involves the transfer of an $\alpha$ particle.
The $\clt$ reaction, as well as $^{12}{\rm C}(^6{\rm Li}, d)^{16}$O, have been used in many indirect measurements of
the $\cago$ cross section (see Refs.\ \cite{BFH78,OHR12} and references therein). This reaction is crucial in
stellar models, since it determines the $^{12}$C and $^{16}$O abundances after helium burning. As astrophysical
energies are much lower than the Coulomb barrier, the corresponding cross sections are too small to be measured
in the laboratory. 

Although many direct measurements have been devoted to the $\cago$ reaction, the extrapolation
down to stellar energies ($\approx 300$ keV) remains uncertain (see Ref.\ \cite{DGW17} for a recent review). Most fits of
the available data are performed within the phenomenological $R$-matrix theory, which involves various parameters of $^{16}$O states.
In particular, the reduced $\alpha$ widths of bound states are proportional to the spectroscopic factors, which can be accessed
by $\alpha$ transfer reactions. Reactions such as $\clt$ therefore provide strong constraints on the $R$-matrix fits.

\begin{figure}[ht]
	\centering
	\includegraphics[width=8.5cm,clip]{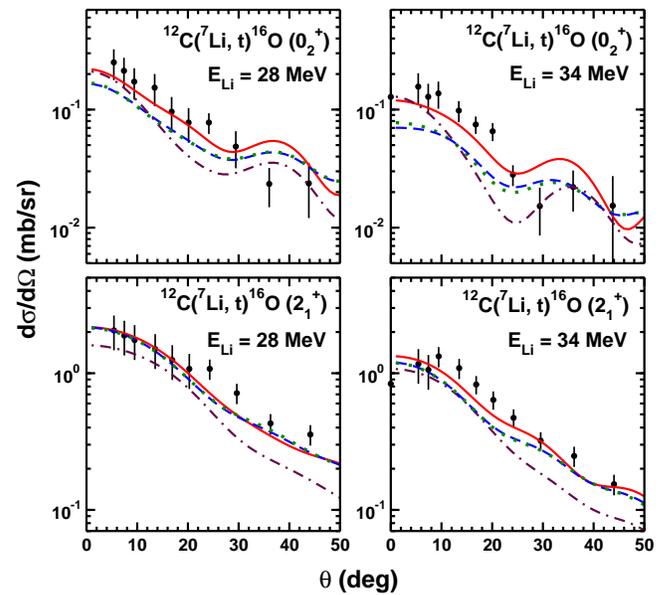}
	\caption{$\clt$ angular distributions  at two different $^7$Li energies. The solid lines correspond to the calculations without the remnant terms. Dashed and dotted lines are obtained by including the remnant terms with the $t-^{12}$C potentials from Ref.\ \cite{LLC07} and Ref.\ \cite{PDM15}, respectively.  
		The dash-dotted lines are obtained with the alternative $\alpha+^{12}$C potentials (see text).
		Experimental data (solid dots) are taken from Ref.\ \cite{OHR12}.  }
	\label{li_t_css}
\end{figure}

We consider the $\alpha$ transfer leading to the $0^+_2$ ($E_x = 6.05$ MeV) and $2^+_1$ ($E_x = 6.92$ MeV) states of $^{16}$O.
Measurements are available for $^7$Li energies of 28 and 34 MeV \cite{OHR12}. The transfer cross sections are presented in
Fig.\ \ref{li_t_css} (solid lines), where we use the spectroscopic factors given in Ref.\ \cite{OHR12} (0.13 for
the $0^+_2$ state and 0.15 for the $2^+_1$ state). The present cross sections are quite similar to the
fits of Ref. \cite{OHR12}, and confirmed by FRESCO calculations (not shown). 

To assess the influence of the remnant term in the DWBA matrix element, we use two different $t+^{12}$C optical potentials
from Refs.\ \cite{LLC07} and \cite{PDM15}. These potentials provide similar elastic-scattering cross sections, and
the shape of the transfer cross section also weakly depends on the core-core potential. The amplitude, however, is 
affected by the presence
of the remnant term. This effect is more significant for the $0^+_2$ state, where the amplitude is changed by about $30\%$. 
This means that the spectroscopic factor should be increased by about $30\%$, compared to the value deduced in Ref.\ \cite{OHR12}

We also want to address the influence of the $\alpha+^{12}$C potentials, associated with $^{16}$O bound states. Following Refs.\
\cite{BFH78,OHR12}, the depths are chosen such that the number of nodes $n$ satisfies the condition $2n+\ell=8$. This
choice seems natural if one considers  pure $\alpha+^{12}$C cluster states, where the four nucleons of the $\alpha$
particle are promoted to the $sd$ shell. Microscopic calculations \cite{Su76c}, however, suggest $2n+\ell=6$. To investigate this
effect, we have repeated the calculations by decreasing the depths of the $\alpha+^{12}$C potentials ($-47.80$ MeV for the
$0^+_2$ state, and $-45.81$ MeV for the $2^+_1$ state). In this way, the $0^+_2$ and $2^+_1$ wave functions present
3 and 2 nodes, respectively. The transfer cross sections are presented in Fig.\ \ref{li_t_css} (dash-dotted lines).
The cross sections are slightly reduced (up to 30\% depending on the angle and energy). As for the effect of the remnant 
term, the choice of the potential does not modify drastically the spectroscopic factors. However these effects suggest
that the error bars on the spectroscopic factors could be underestimated, owing to uncertainties in the model.

\begin{figure}[ht]
	\centering
	\includegraphics[width=7.5cm,clip]{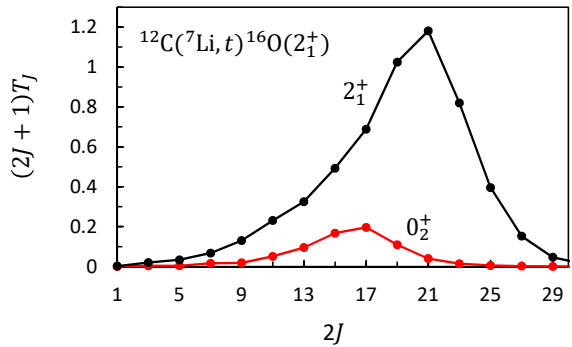}
	\caption{Coefficients $(2J+1)T_J$ \ref{eq15c} for the $\clt$ 
		reaction at $E_{Li} =  28$ MeV as a function of $J$ and for two $^{16}$O states. The lines are guide to the eye. }
	\label{dp_J_cli}
\end{figure}

In Fig.\ \ref{dp_J_cli}, we present the values of $(2J+1)T_J$ for $E_{\rm Li}=28$ MeV. For the dominant $2^+_1$ state,
the maximum is located near $J\approx 21/2$, whereas it is shifted down to around $J\approx 17/2$ for the $0^+_2$
state. Partial waves above $J\gtrsim 29/2$ (i.e. for $L_A \gtrsim 14$) play a negligible role.

\subsection{Test of the peripherality}
DWBA calculations are widely used to determine ANCs from transfer data \cite{TBC14}. The ANC of a bound state in
the residual nucleus $B$ is defined by
\begin{align}
u^{I_B}_{\ell_B}(r_B)\longrightarrow C^{I_B}_{\ell_B}\, W_{-\eta_B,\ell_B+1/2}(2\kappa_Br_B),
\label{eq31b}
\end{align}
where $\eta_B$ and $\kappa_B$ are the Sommerfeld parameter and wave number, and $W_{a,b}(x)$ is the Whittaker function.
In most cases, it is assumed
that the transfer process is essentially peripheral, and therefore probes the long-range part of the wave functions. This problem
has been addressed, for example, in Ref.\ \cite{PNM07}. Assessing the peripherality of a transfer reaction, however,
is not trivial. The main reason is that the transfer kernel (\ref{eq14}) explicitly depends on the relative
coordinates between the colliding nuclei ($R$ and $R'$), whereas the peripheral character is associated with the internal
coordinates $r_A$ and $r_B$ (see Fig. \ref{fig1}).

To analyze the peripheral nature of a transfer reaction, we define a modified kernel as
\begin{align}
\tilde{K}^{J\pi}_{\alpha \beta}(\rmin,R,R') &= K^{J\pi}_{\alpha \beta}(R,R') {\rm \ for \ } \rmin\leq r_A {\rm \ or \ } r_B \nonumber \\
&=0  {\rm \ for \ } \rmin > r_A {\rm \ or \ } r_B
\label{eq32}
\end{align}
where $\rmin$ is a cutoff radius on the internal coordinates $r_A$ or $r_B$. This definition provides a modified
scattering matrix as
\begin{align}
\tilde{U}^{J\pi}_{\alpha \beta}(\rmin)= -\frac{i}{\hbar}\int& 
\chi^{J\pi}_{L_A}(R) \tilde{K}^{J\pi}_{\alpha \beta}(\rmin,R,R') \chi^{J\pi}_{L_B}(R')\nonumber \\
&\times RR'dRdR'.
\label{eq33}
\end{align}
Consequently we have
\begin{align}
&\tilde{U}^{J\pi}_{\alpha \beta}(0) = U^{J\pi}_{\alpha \beta} \nonumber \\
&\tilde{U}^{J\pi}_{\alpha \beta}(\infty) = 0.
\label{eq34}
\end{align}
The cutoff radius $\rmin$ can be applied, either to the internal coordinate of the projectile ($r_A$) or of the residual
nucleus ($r_B$). For a peripheral process, one expects $\tilde{U}^{J\pi}_{\alpha \beta}(\rmin)$ to have significant values for
large $\rmin$. In contrast, if $\tilde{U}^{J\pi}_{\alpha \beta}(\rmin)$ tends rapidly to zero, the process can be considered as 
internal. Notice that the peripheral nature depends on the angular momentum $J$.

\begin{figure}[ht]
	\centering
	\includegraphics[width=7.5cm,clip]{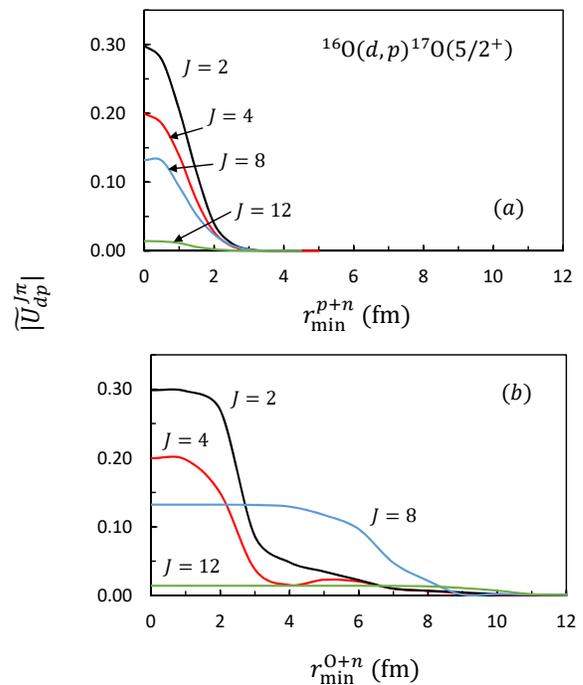}
	\caption{Modulus of the modified scattering matrix $\tilde{U}^{J\pi}$ (\ref{eq33}) for the $\odp$(g.s.) reaction at $E_d =  25.4$ MeV, and
		for different $J$ values. 
		The minimum distance $\rmin$ corresponds to the $p+n$ coordinate in panel (a) and to the $^{16}$O+n coordinate in
		panel (b). }
	\label{fig_peri_odp_u}
\end{figure}

As a first test, we consider the $\odp$ reaction to the $5/2^+$ ground state (we have selected $I_B=2, L_B=\vert J-2 \vert$, but
similar conclusions are obtained for other quantum numbers). In Fig.\ \ref{fig_peri_odp_u}, we present the modified
scattering matrices (\ref{eq33}) as a function of $\rmin$ in the deuteron (a) and in $^{17}$O (b). 
As it is well known for $(d,p)$ reactions, the transfer process is sensitive to short $p+n$ distances only (Fig.\ \ref{fig_peri_odp_u}a). Above 2 fm,
the contribution to the scattering matrix is negligible. This result justifies the zero-range approximation which is
often used in the literature for $(d,p)$ and $(d,n)$ reactions. The situation is different for the $n+^{16}$O distance
(Fig.\ \ref{fig_peri_odp_u}b). For large angular momenta ($J\gtrsim 8$) the integral is mostly sensitive to large distances, 
and the reaction can be considered as peripheral. However, low $J$ values ($J\lesssim 8$) mostly depend on the internal contribution.

\begin{figure}[ht]
	\centering
	\includegraphics[width=7.5cm,clip]{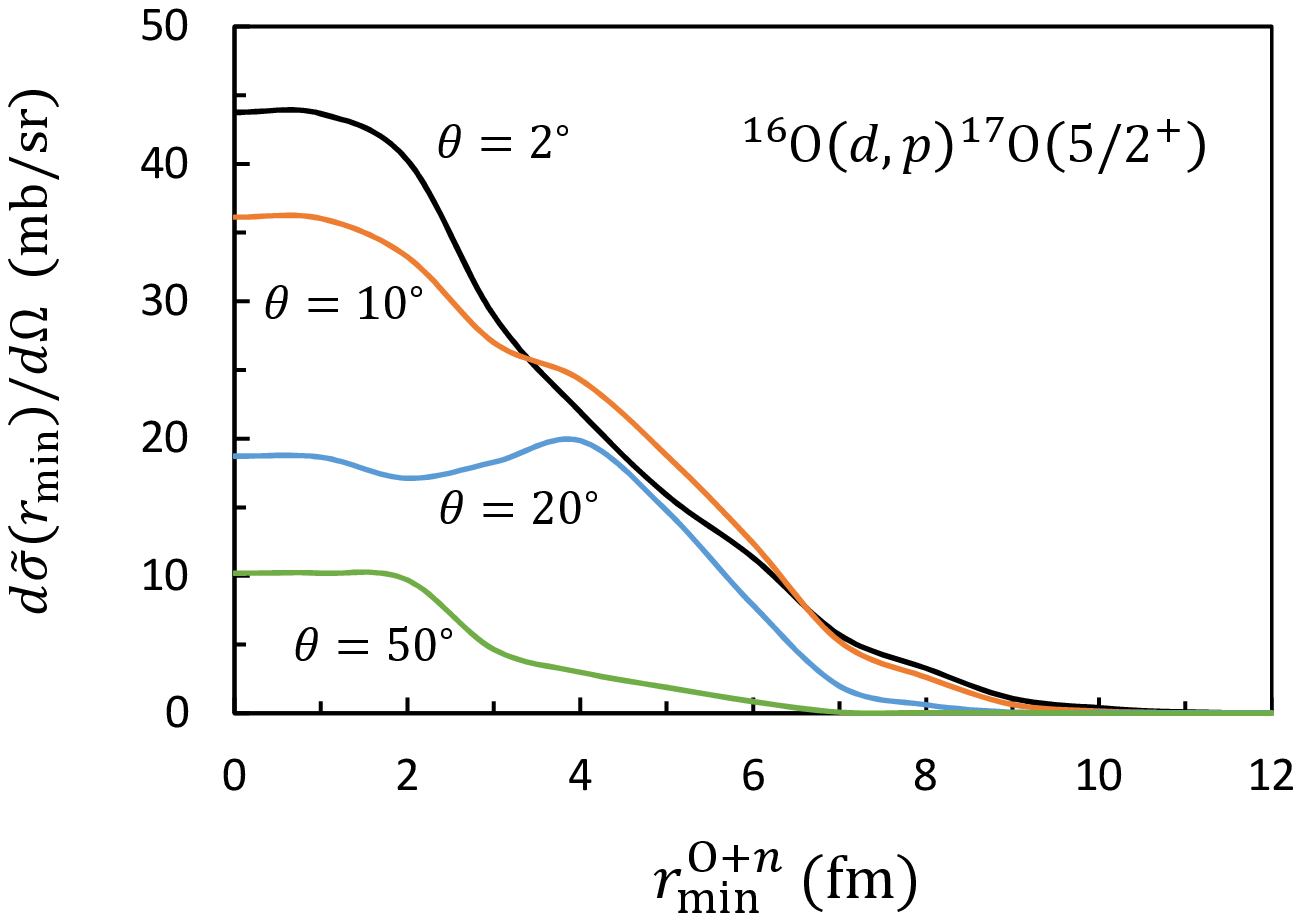}
	\caption{Modified scattering cross sections $d\tilde{\sigma}(\rmin)/d\Omega$ for the $\odp$(g.s.) reaction at $E_d =  25.4$ MeV, and for different scattering angles. }
	\label{fig_peri_odp_sig}
\end{figure}

The different behaviour of the $J$ values suggests that the peripheral nature of the cross section depends on the angle. 
This is confirmed in Fig.\ \ref{fig_peri_odp_sig}, where we plot the modified cross sections
$d\tilde{\sigma}(\rmin)/d\Omega$, computed with the scattering matrices (\ref{eq33}). At small angles, the cross section can be 
considered as essentially peripheral. However, the situation is different when the angle increases. This property is
expected to be valid in other systems, and suggests that the determination of ANC should be limited to small angles.

\begin{figure}[ht]
	\centering
	\includegraphics[width=7.5cm,clip]{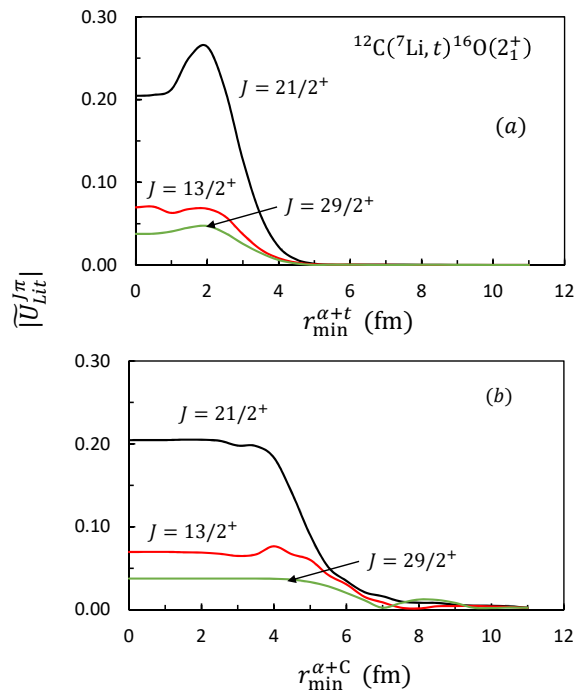}
	\caption{Modulus of the modified scattering matrix $\tilde{U}^{J\pi}$ (\ref{eq33}) for the $\clt (2^+_1)$ reaction at $E_{\rm Li} =  28$ MeV. 
		The minimum distance $\rmin$ corresponds to the $\alpha+t$ coordinate in panel (a) and to the $\alpha+^{12}$C coordinate in
		panel (b). }
	\label{fig_peri_cli_u}
\end{figure}

\begin{figure}[ht]
	\centering
	\includegraphics[width=7.5cm,clip]{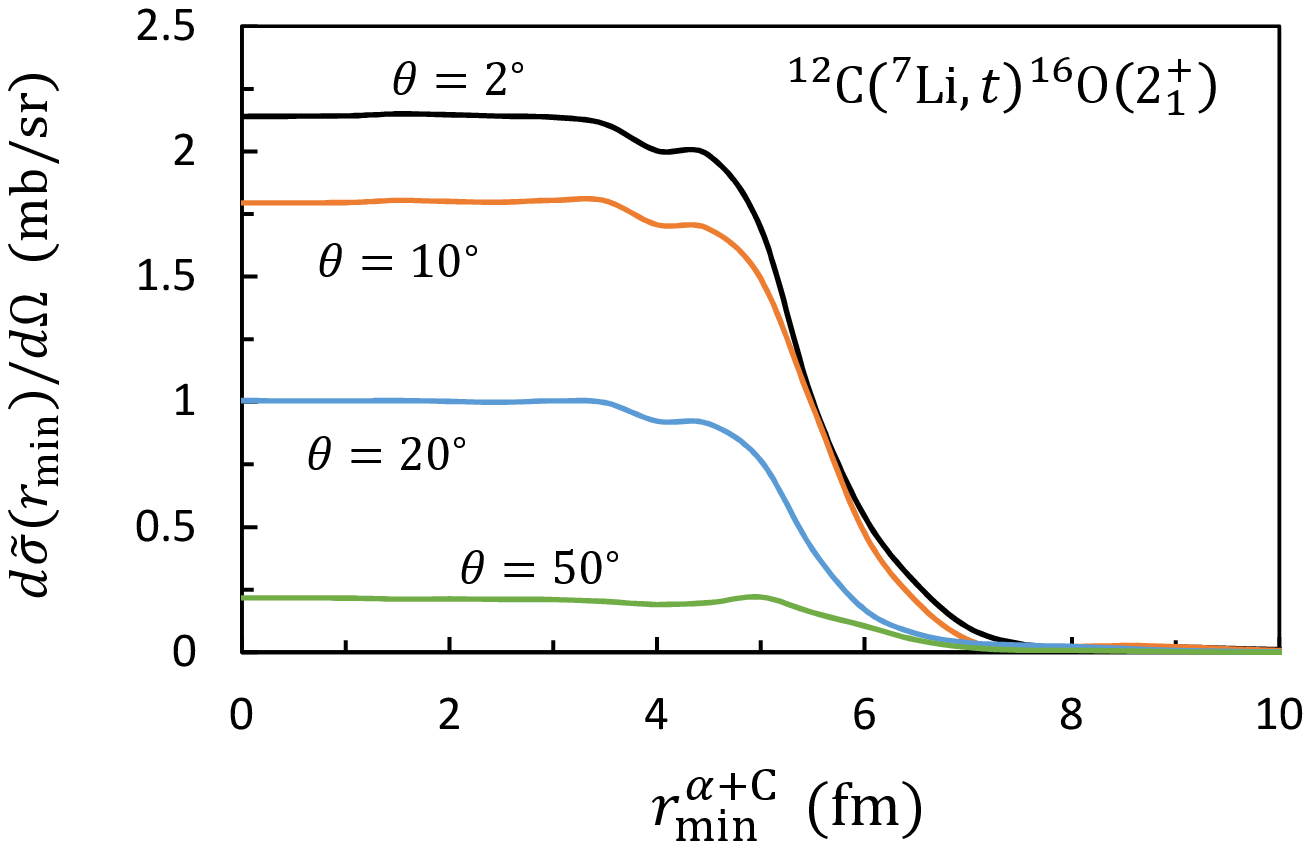}
	\caption{Modified scattering cross sections $d\tilde{\sigma}(\rmin)/d\Omega$ for the $\clt (2^+_1)$ reaction at $E_{\rm Li} =  28$ MeV, 
		and for different scattering angles. }
	\label{fig_peri_cli_sig}
\end{figure}

Figures \ref{fig_peri_cli_u} and \ref{fig_peri_cli_sig} represent the same quantities for the $\clt (2^+_1)$ reaction.
Figures \ref{fig_peri_cli_u} illustrates the bahaviour of the scattering matrix for typical $J$ values. In that case, the zero-range
approximation would not be accurate since the dependence on $\rmin$ in the $\alpha+t$ motion is important.
From Fig.\ \ref{fig_peri_cli_u}(b), we conclude that the transfer process is not sensitive to the $\alpha+^{12}$C coordinate
below $\approx 4$ fm. The modified cross sections presented in Fig.\ \ref{fig_peri_cli_sig} confirm this property, which is
mainly due to the weak binding energy of the $2^+_1$ state ($-0.24$ MeV). The wave function in (\ref{eq31b}) presents a
slow decrease, and the transfer process is essentially determined from the large distances.

\section{Conclusions}
\label{sec4}
In this work, we have applied the $R$-matrix method to direct transfer reactions. Using a Lagrange mesh as
basis for the internal wave functions provides an efficient tool to compute the various matrix elements. With the $\odp$
and $\clt$ reactions, we have considered typical neutron and $\alpha$-transfer processes. We have shown that this
framework requires a relatively small number of points ($\sim 40-50$), much smaller than in finite-difference
methods. Of course, computer times do not represent a critical issue for simple calculations, as for those associated with
standard DWBA calculations. However, large-scale scattering calculations \cite{CGM17,De18} are more and more demanding 
in terms of computing capabilities, and the present method significantly reduces computer times.

In this exploratory work, we did not aim at fitting data. The potentials were taken from the literature, and were
used to assess the accuracy of the method. We also analyzed the peripherality of the $\odp$ and $\clt$ reactions,
as typical examples. This is done by defining modified transfer kernels, which are set to zero if the internal 
coordinates (in the projectile or in the residual nucleus) is not in a given interval. We have shown that the external contribution to
the scattering matrix depends on angular momentum: small $J$ values are essentially internal, whereas large $J$ values
are more peripheral. Consequently, the peripheral nature of transfer reactions is sensitive to the energy and angle.

In addition to the simplicity of the $R$-matrix, this method also permits to deal with non-local potentials \cite{HRB02}.
Going beyond the DWBA leads to non-local potentials \cite{OII69,CV76,Th82}. In this way, the non-orthogonality
of the entrance and exit channels is treated exactly. In other words, and in contrast with the DWBA, elastic scattering is
modified by the coupling to transfer channels. Although this effect is expected to be small in stable nuclei \cite{Sa83}, it might 
be more important in reactions involving exotic nuclei. 

The $R$-matrix formalism could also be applied to the 
source method, an alternative theory where the transfer scattering matrix is obtained from an inhomogeneous 
equation  \cite{AG69}. Although most $R$-matrix calculations are performed for homogeneous equations, the extension is straightforward.
The present formalism therefore opens several perspectives in future calculations of transfer reactions.

\section*{Acknowledgments}
We thank A. Moro for many helpful discussions.
This work was supported by the Fonds de la Recherche Scientifique - FNRS under Grant Numbers 4.45.10.08 and J.0049.19.
Computational resources have been provided by the Consortium des Équipements de Calcul Intensif (CÉCI), funded by the Fonds de la Recherche Scientifique de Belgique (F.R.S.-FNRS) under Grant No. 2.5020.11 and by the Walloon Region.
P. D. is Directeur de Recherches of F.R.S.-FNRS, Belgium. 

\onecolumngrid
\appendix
\section{}
\label{appendix}
In this Appendix, we give detail about some technical aspects of the DWBA. All coordinates must be expressed
in terms of $(\pmb{R},\pmb{R}')$. Let us define 
\begin{align}
\pmb{r}_A= &\alpha \pmb{R}+ \beta \pmb{R}' \nonumber \\
\pmb{r}_B= &\gamma \pmb{R}+ \delta \pmb{R}' \nonumber \\
\pmb{r}_{ab}= &\omega \pmb{R}+ \nu \pmb{R}' .
\label{eqa1}
\end{align}
A simple calculation provides
\begin{align}
& \alpha= -\frac{bA}{v(A+b)}, \ \beta=\frac{AB}{v(A+b)} \nonumber \\
& \gamma= \frac{AB}{v(A+b)}, \ \delta=-\frac{aB}{v(A+b)} \nonumber \\
& \omega= \frac{b}{A+b}, \ \nu=\frac{B}{A+b}.
\label{eqa2}
\end{align}

The Jacobian is therefore defined by a $6\times 6$ matrix from
\begin{align}
d\pmb{r}_A d\pmb{R}=d\pmb{r}_B d\pmb{R}'={\cal J}d\pmb{R}d\pmb{R}',
\label{eqa3}
\end{align}
with
\begin{align}
{\cal J}=\beta^3.
\label{eqa4}
\end{align}

Let us outline the calculation of the transfer kernel (\ref{eq14}), which represents integrals over
the angles $\Omega$ and $\Omega'$. The main difficulty is that these angles show up through coordinates
$(\pmb{r}_A,\pmb{r}_B,\pmb{r}_{ab})$, as shown by Eq.\ (\ref{eqa1}). To simplify the presentation, we 
assume here that the spins of $a$ and $v$ are zero. The first step is to expand the potentials and
radial wave functions as
\begin{align}
\frac{u^{I_A}_{\ell_A}(r_A)}{r_A^{\ell_A+1}}
\frac{u^{I_B}_{\ell_B}(r_B)}{r_B^{\ell_B+1}}
\bigl(V_{av}(\pmb{r}_A)+V_{rem}(\pmb{r}_{ab},\pmb{R}')\bigr)=
\sum_K g^K_{\alpha\beta}(R,R')\bigl[Y_K(\Omega)\otimes Y_K(\Omega')\bigr]^0,
\label{eqa5}
\end{align}
which is performed by a numerical quadrature over the angle between $\pmb{R}$ and $\pmb{R}'$.

To proceed further, we use the expansion (assuming $\pmb{S}=\alpha \pmb{r}_1+\beta \pmb{r}_2$)
\begin{align}
S^L Y_L^M (\Omega_S)=\sum_k C_L^k(\alpha r_1)^k(\beta r_2)^{L-k}
\bigl[Y_k(\Omega_1)\otimes Y_{L-k}(\Omega_2)\bigr]^{LM},
\label{eqa6}
\end{align}
with
\begin{align}
C_L^k=\biggl(\frac{4\pi (2L+1)!}{(2k+1)!(2L-2k+1)!}\biggr)^{1/2}.
\end{align}With expansions (\ref{eqa5}) and (\ref{eqa6}), the transfer kernel can be written as
\begin{align}
K^{J\pi}_{\alpha \beta}(R,R')=\sum_{k_1 k_2 K}
C^{k_1}_{\ell_A} C^{k_2}_{\ell_B} 
F^{J}_{\alpha \beta,k_1 k_2 K}
I_{\alpha \beta,k_1 k_2 K}(R,R'),
\end{align}
where functions $I_{\alpha \beta,k_1 k_2 K}(R,R')$ are given by
\begin{align}
I_{\alpha \beta,k_1 k_2 K}(R,R')=
(\alpha R)^{k_1}(\beta R')^{\ell_A-k_1}
(\gamma R)^{k_2}(\delta R')^{\ell_B-k_2}
g^{K}_{\alpha \beta}(R,R')
\end{align}
and coefficients $F^{J\pi}_{\alpha \beta,k_1 k_2 K}$ by
\begin{align}
F^{J\pi}_{\alpha \beta,k_1 k_2 K}=\langle \biggl[ Y_{L_A}(\Omega)\otimes
\bigl[Y_{k_1}(\Omega)\otimes Y_{\ell_A-k_1}(\Omega')\bigr]^{\ell_A}\biggr]^J \vert
\bigl[Y_K(\Omega)\otimes Y_K(\Omega')\bigr]^0 \vert
\biggl[ Y_{L_B}(\Omega')\otimes
\bigl[Y_{k_2}(\Omega)\otimes Y_{\ell_B-k_2}(\Omega')\bigr]^{\ell_B}\biggr]^J
\rangle
\end{align}
The analytical calculation of these coefficients requires some algebra to modify the order of angular-momentum
couplings, and involves $6j$ coefficients. When particles $a$ and $v$ have a spin, further angular-momentum recoupling
is necessary. This calculation does not raise particular difficulties. It is developed in more detail, for
example, in Ref.\ \cite{Sa83}.

%

\end{document}